\documentclass[modern]{aastex631}
\received{April 1, 2021}
\shorttitle{I'll Finish It This Week}
\shortauthors{Brauer}
\graphicspath{{./}{figures/}}
\begin{document}

\title{``I'll Finish It This Week" And Other Lies}

\author{Kaley Brauer}
\affiliation{Department of Physics and Kavli Institute for Astrophysics and Space Research, Massachusetts Institute of Technology, Cambridge, MA 02139, USA}



\begin{abstract}

A small group of postdocs, graduate students, and undergraduates inadvertently formed a longitudinal study contrasting expected productivity levels with \textit{actual} productivity levels. Over the last nine months, our group self-reported 559 tasks, dates, and completion times  -- expected and actual. Here, I show which types of tasks we are the worst at completing in the originally planned amount of time (spoiler: coding and writing tasks), whether more senior researchers have more accurate expectations (spoiler: not much), and whether our expectations improve with time (spoiler: only a little).

\end{abstract}


% \keywords{x}


\section{Introduction} \label{sec:intro}

How many times have you sworn that this is the week you finally fix that bug in your code? How many times have you promised that this is the month you finally finish that paper draft? And how many times does it actually happen?

I am like you. I lie to myself constantly and copy-paste my to do list from one week to the next. But in the summer of 2020, a small group of postdocs, graduate students (including myself), and undergraduates came together to motivate and support each other on their weekly goals. The idea was simple: a weekly check-in where we share the tasks we plan to do that week and how long we expect each of them to take. Then, we report back on how long the tasks \textit{actually} take. Without intending, we created a data set contrasting our expectations and our realities.

%A global pandemic is not an opportune time to try to get your life organized. 
The last year has emphasized that there are far more important things than how productive you are in a week. At the same time, for us, meeting weekly provided a bit of needed structure in the chaos of our year and planning out our days let us pretend to embrace some sort of work-life balance.
So, googling ``\textit{time management??}" and ``\textit{what is a SMART goal anyway}", we set off on a nine-month journey to try to understand our own limits and expectations.

Here, I present how we did and what we learned. Maybe you'll find insight too.

\section{The Data} \label{sec:data}

The data presented here covers the period from June 22, 2020 until present. It includes 559 self-reported tasks, dates of completion (planned and actual), and completion times in terms of active working hours (planned and actual). The data does not include tasks that are scheduled for a specific amount of time on a specific day (e.g., meetings or lectures). It only includes tasks for which the actual completion time is unknown a priori, henceforth called ``unstructured'' work.

Every task was intended to be a specific, realistic goal that we would complete that week. The tasks are split into eight categories:
\begin{enumerate}
\item Coding: any coding task for research or schoolwork, e.g., analysis in python.
\item Writing: any writing- or editing-focused task, e.g., working on a paper draft.
\item Reading: any reading-focused task, e.g., reading a journal article.
\item Administrative: any task related to running a research group or department, e.g., organizing meetings.
\item Talk Prep: writing or practicing a talk or poster presentation.
\item Service: volunteer work, e.g., organizing outreach activities.
\item Problem Set: homework for a class that is not coding- or reading-focused.
\item Other
\end{enumerate}

The data includes tasks reported by two postdocs, four graduate students, and two undergraduate students. The participants released their data for this paper.
%The participants released their data to be analyzed and shared in aggregate.

\section{The Results}

Overall, we complete 53\% of our weekly tasks in the originally planned week. The actual number of hours required to complete a task is, on average, 1.7x as many hours as expected (with a median multiplier of 1.4x). Shockingly accurate, to be honest.

\subsection{Different Types of Tasks}

\begin{figure*}[htb!]
\center
\gridline{
	\fig{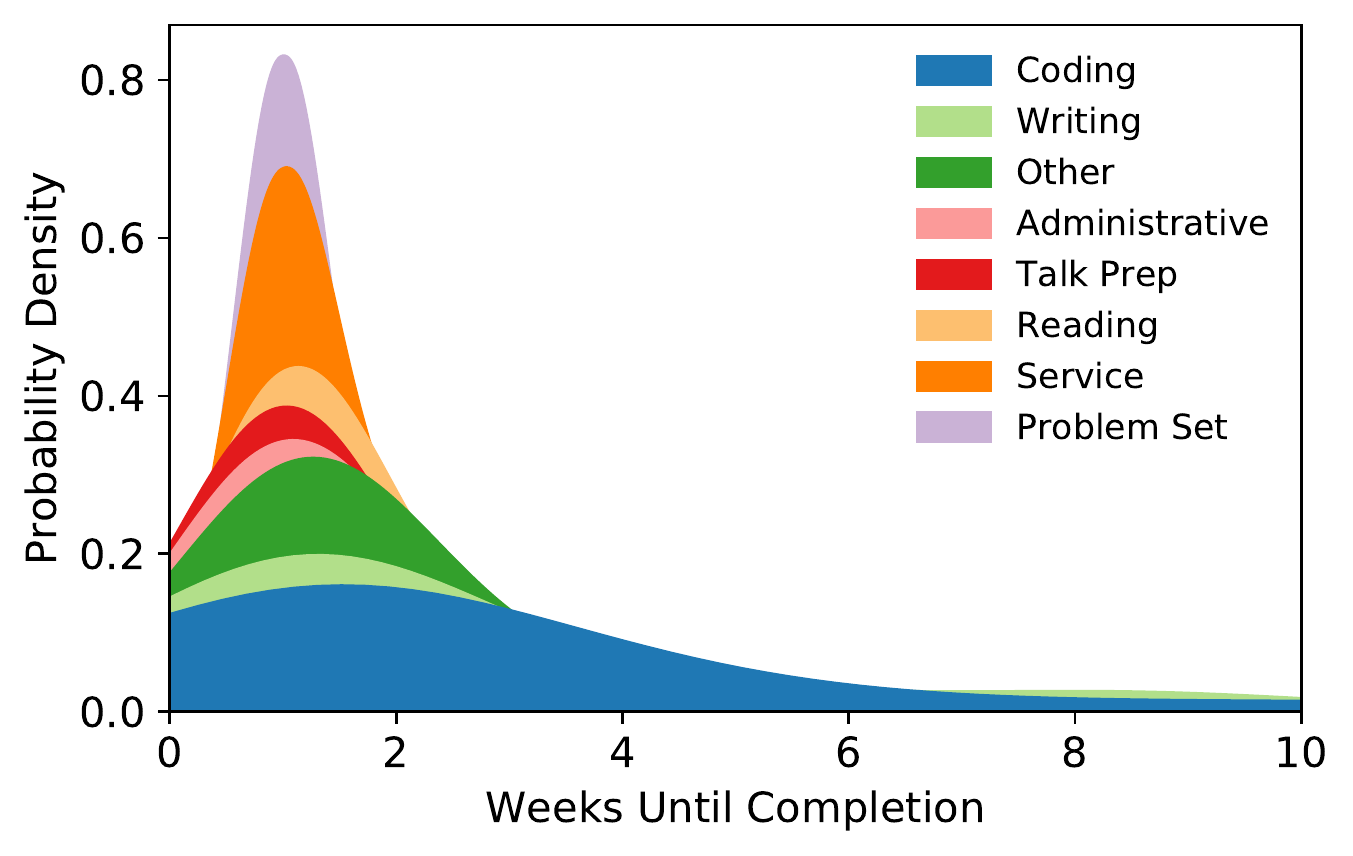}{0.65\textwidth}{(a) How many weeks pass before different types of tasks are completed. All tasks are expected to be completed in one week.}}
\gridline{
	\fig{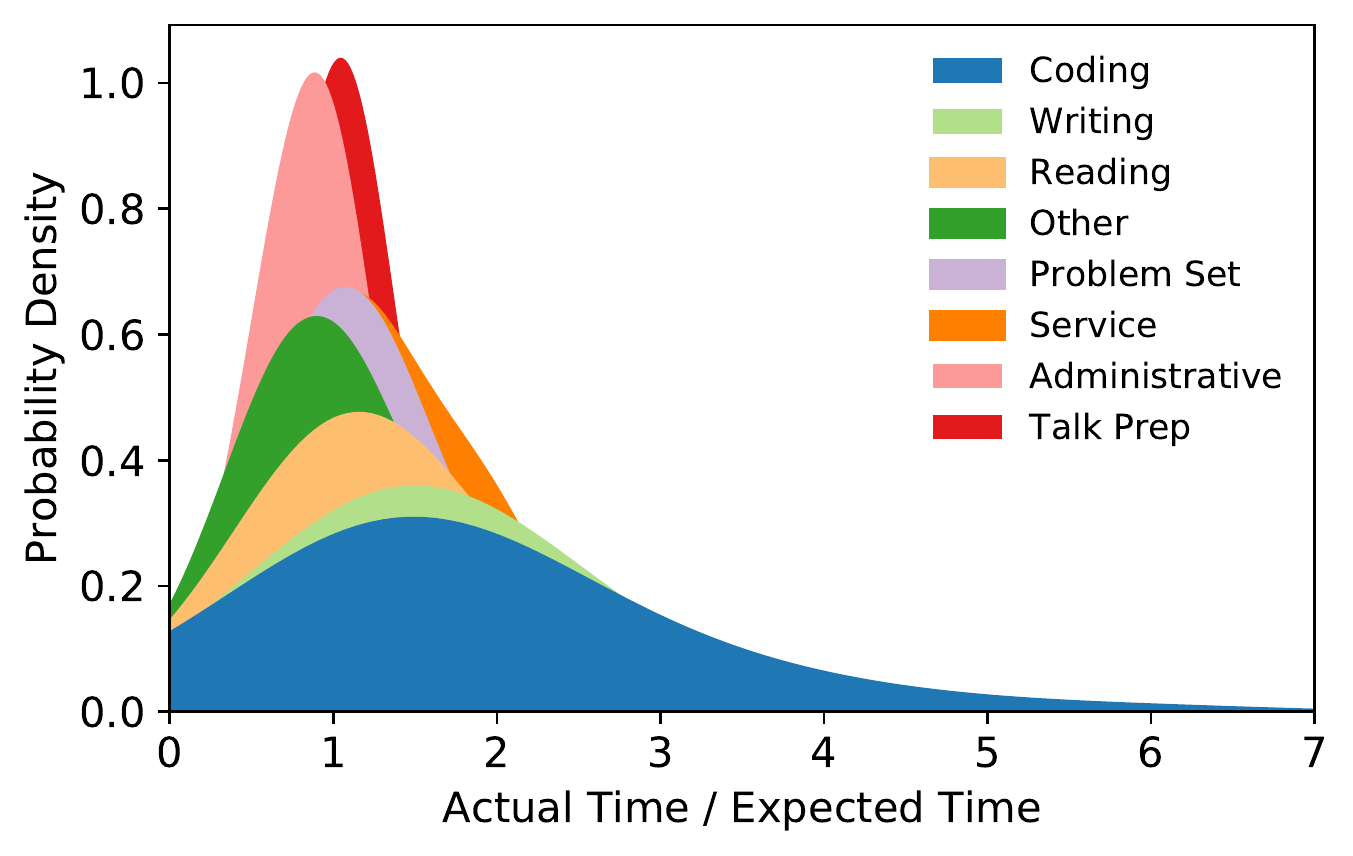}{0.65\textwidth}{(b) How our estimated hours compare to the actual number of active hours to complete different tasks.}}
\caption{Gaussian kernel density estimations showing how quickly different tasks are completed. If our expected productivity levels were perfectly accurate, the probability densities of both plots would be delta functions centered on $x=1$.
\label{fig:types}}
\end{figure*}

Which tasks do we estimate accurately, and which ones can be train wrecks? Answer: if a train goes off the rails, it is probably Coding or Writing. Figure \ref{fig:types} shows estimated probability densities for how quickly we complete different types of tasks. Histograms of the data are shown in the Appendix in Figure \ref{fig:hist}.

Figure \ref{fig:types}a shows how many weeks pass before a task is completed. In theory, every task should be completed in one week. In reality, Coding and Writing tasks can take \textit{much} longer. The worst example from this data is a Coding task that took 31 weeks to complete (note: the data only covers 35 weeks, so this is nearly as bad as possible for a completed task). Still, the average weeks-until-completion for Coding and Writing tasks are both within a month (2.8 and 3.4 weeks, respectively) and only about 10\% took over two months to complete.

While Coding and Writing tasks are a toss up (could be completed this week, could be completed in 8 months), Problem Set and Service tasks are virtually always completed when expected. These tasks also virtually always have deadlines (the problem set is due or the volunteer event is about to occur), which is not a coincidence.

Figure \ref{fig:types}b shows how accurate we are at estimating the number of active hours different tasks will require. Once again, our expectations are the worst for Coding and Writing tasks. Two Coding tasks and one Writing task took over 10x as long as expected. The most common Actual/Expected time for Coding and Writing tasks is only 1.5x, however. Other types of tasks are estimated with approximate accuracy.

\subsection{Different Career Levels}

Do more senior researchers (e.g., postdocs) have more accurate expectations for their work than more junior researchers (e.g., undergraduate or graduate students)? Answer: not really, but this is likely related to the types of tasks done by the different groups.

\begin{figure*}[tbh!]
\gridline{
	\fig{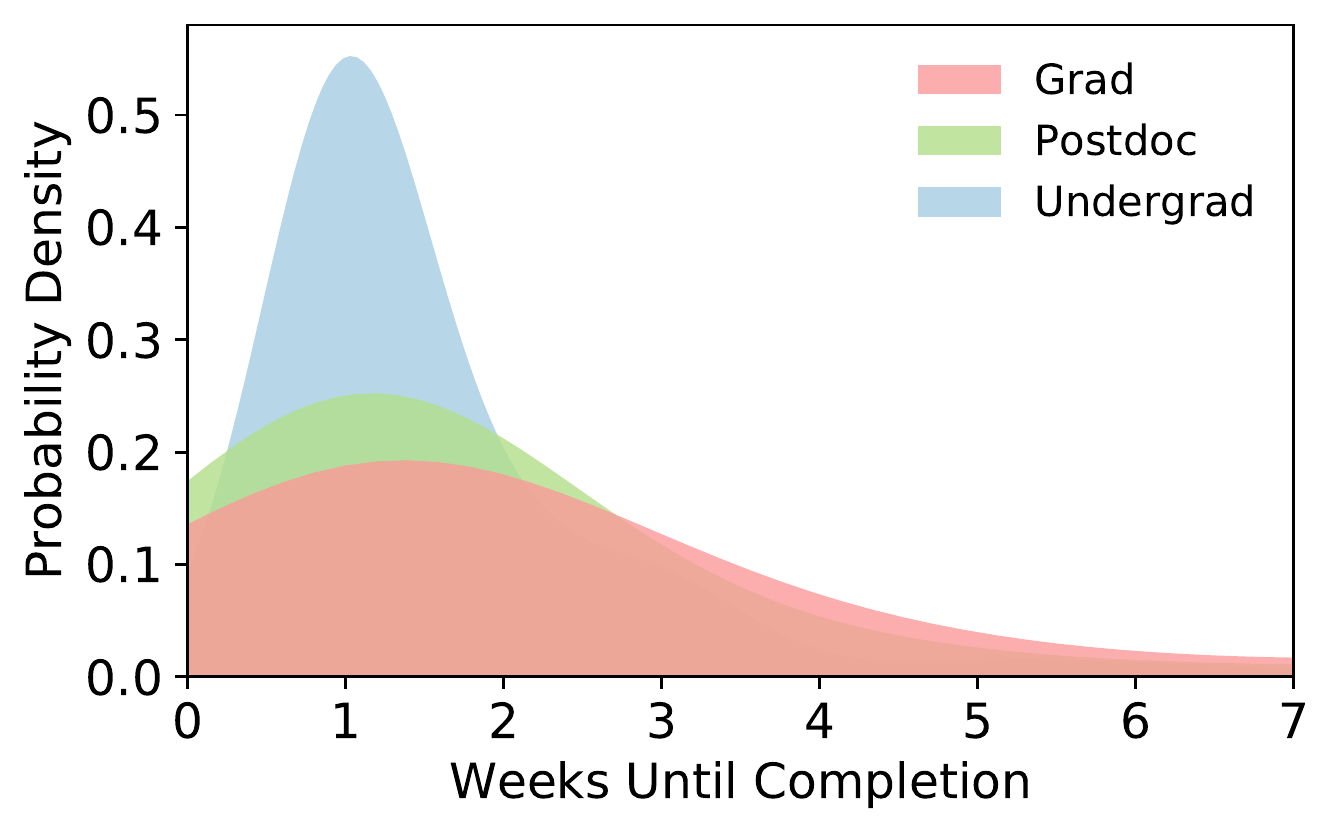}{0.5\textwidth}{(a) How many weeks pass before different career levels complete their tasks. All tasks are expected to be completed in one week.} 
	\fig{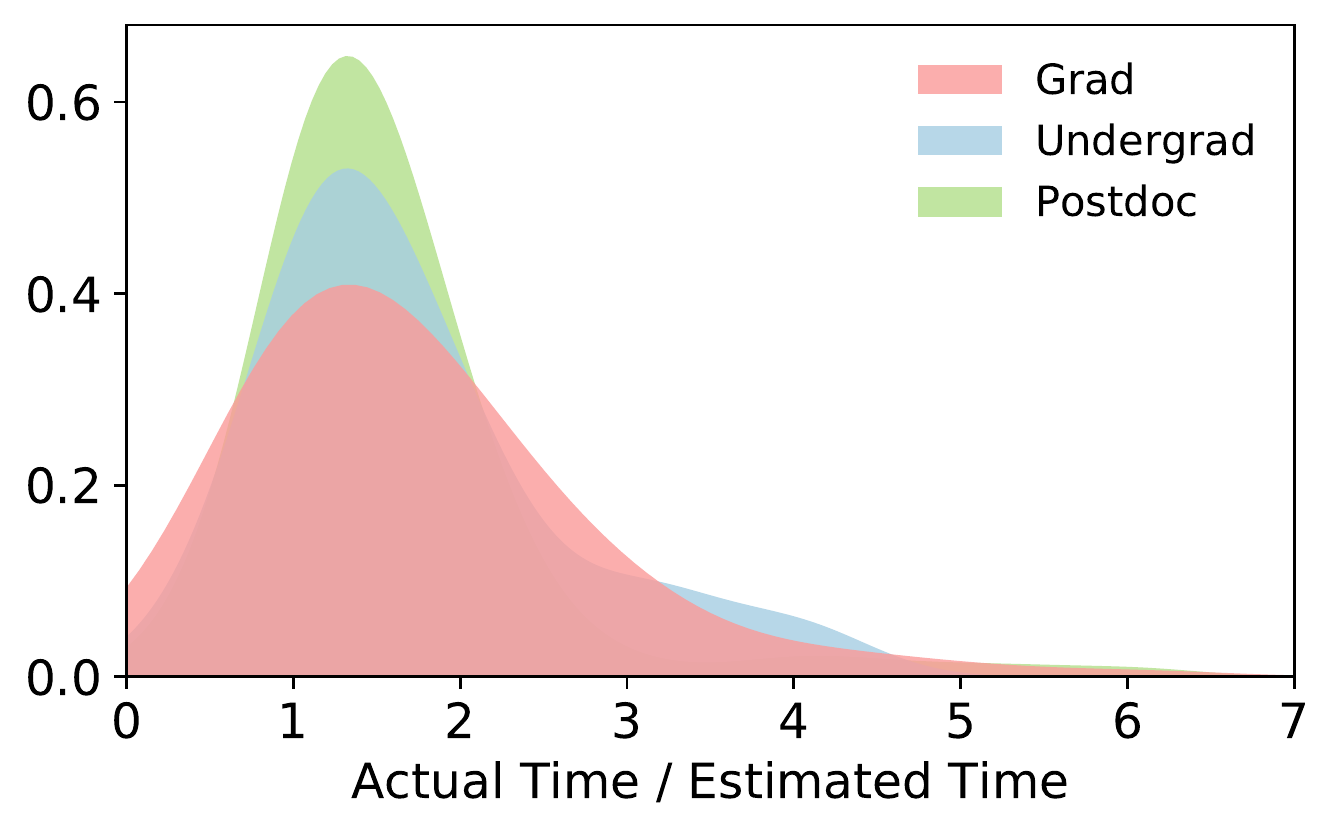}{0.5\textwidth}{(b) How accurately different career levels are able to estimate the number of active hours needed to complete their tasks.}}
\caption{Gaussian kernel density estimations showing how quickly tasks are completed by researchers at varying career levels.
\label{fig:prof}}
\end{figure*}

Figure \ref{fig:prof} shows estimated probability densities for how quickly different groups complete their tasks. As seen in Figure \ref{fig:prof}a, the undergraduates in the study are most likely to complete their tasks in the planned week. The tasks planned by undergraduates are also more likely to have short deadlines, though (e.g., classwork).

Figure \ref{fig:prof}b shows that all three groups are most likely to take about 1.4x as long as they expect to finish any random task. Grad students have the heaviest tail on their distribution, though, and thus are more likely than the other groups to spend longer than expected on a task despite being more senior than undergrads.

\begin{figure*}[!bth]
\center
\includegraphics[width=0.75\textwidth]{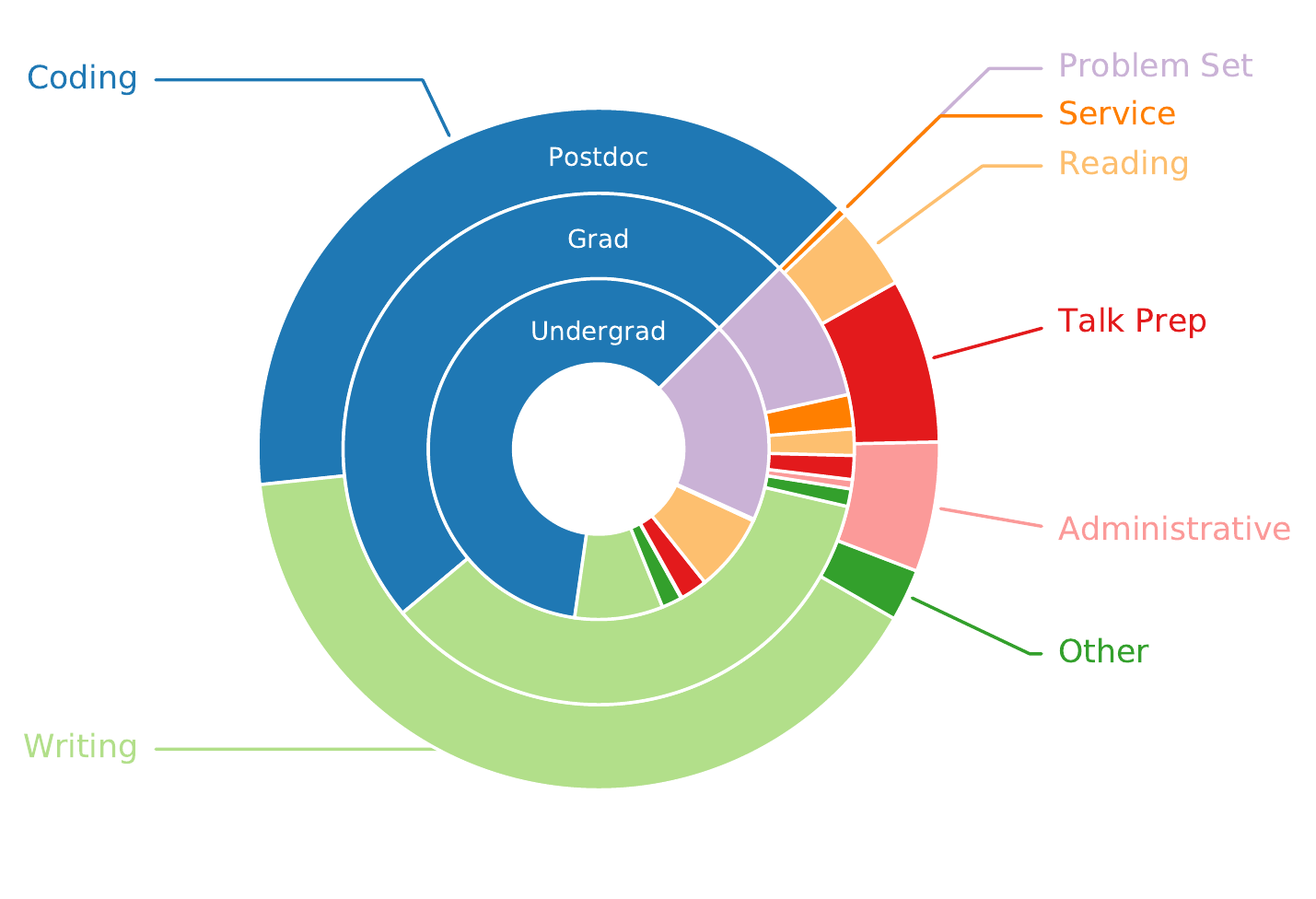}
\caption{Pie charts showing the fraction of our unstructured work time spent doing different types of tasks. Meetings and lectures were not included in the data (these are ``structured tasks''), but that is probably for the best because I don't want to know how much time we all spent sitting in Zoom meetings.
\label{fig:proftimes}}
\end{figure*}

Figure \ref{fig:proftimes} shows a breakdown of how we spent our unstructured work time over the last nine months. As one might expect, time spent doing Problem Sets decreases with career level while time spent Writing and Administrating increases. I would love to see the Faculty breakdown too, but I am not positive they have any unstructured time.

\subsection{Improvement Over Time}

Did we get better at planning our weeks over time? Answer: ...a little! Looking only at data from the four members who participated for the full nine months (one postdoc, two grad students, one undergraduate), in the first three months, we completed 47\% of our weekly tasks in the originally specified week. In the middle three months, we completed 59\% of tasks on time! And in the last three months, we completed 51\% of tasks on time.

\begin{figure}[htb!]
\center
\includegraphics[width=0.5\textwidth]{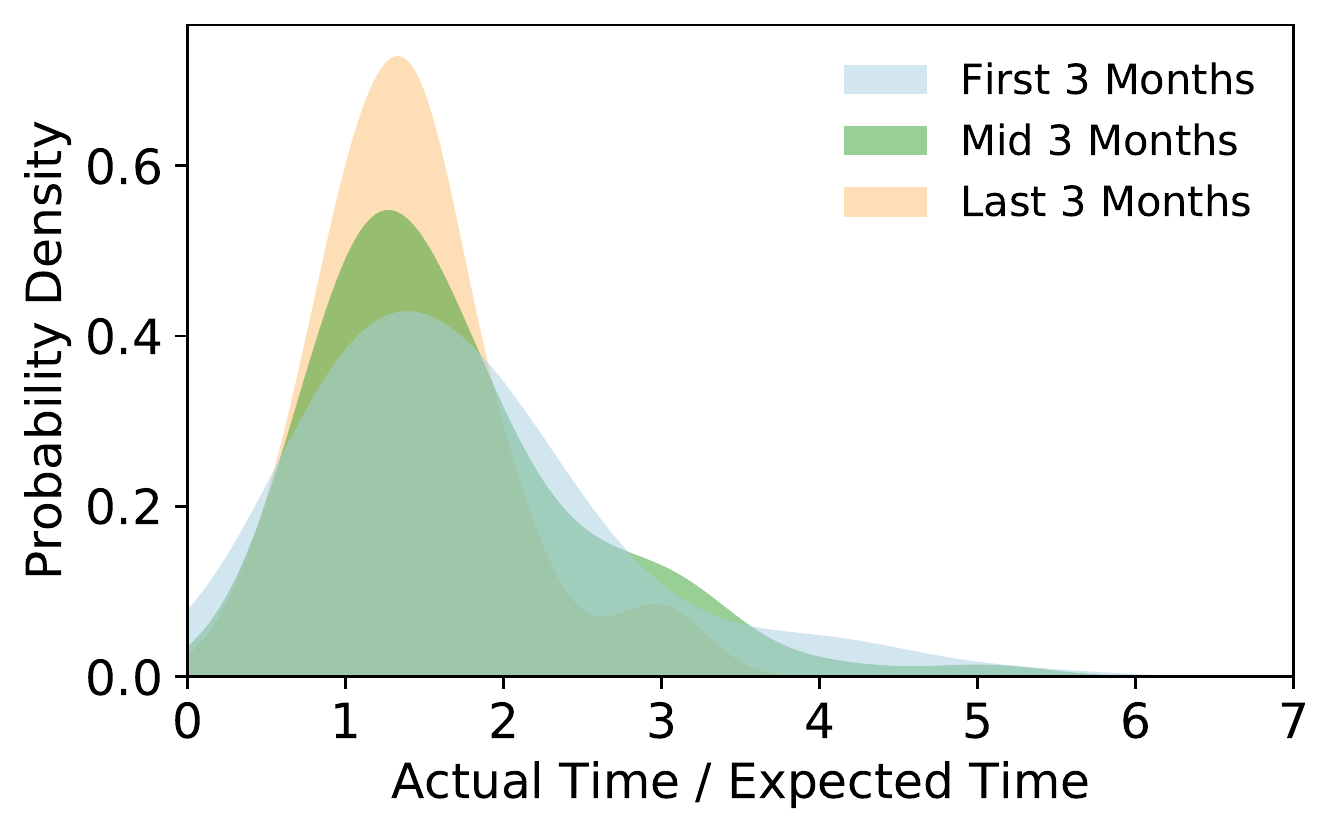}
\caption{Probability density of how our estimated hours compared to the actual number of active hours to complete a task at the beginning, middle, and end of the study. We became slightly more accurate in our expectations over time.
\label{fig:overtime}}
\end{figure}

While the percentage of tasks we completed in the planned week did not strictly improve, our ability to estimate how many active hours of work are needed for a task improved a little bit as the months went on. Figure \ref{fig:overtime} shows how accurate our hour estimates were during different parts of the study. There is a trend towards more accurate estimates over time!

\subsection{The Tasks We Didn't Finish}

As of the end of this study, 96 of the 559 planned tasks had no reported completion date (about half of which are coding tasks). This means that up to 17\% of our tasks were abandoned or still incomplete. Not bad.

\section{The Takeaways}

So, what did we learn?

\begin{enumerate}
    \item If you are coding or writing, multiply the expected time of any specific task by $\sim1.5$x (the real completion time will now probably still be 1.5x longer than your new expected time, but you tried).
    \item Don't expect more senior people to necessarily be better at predicting completion times.
    \item Don't expect to quickly improve your expectations of what you can do each week.
    \item Tasks with deadlines are much more likely to be done quickly (though at what cost).
\end{enumerate}

Additionally, some members realized that time-based goals (``work on research for 5 hours'') can work better than results-based goals (``solve research problem in 5 hours''). Find what works for you. Alternatively, if you just want to pump up your fraction of completed tasks, you could always pad your to do list with more short, ``easy'' tasks. Or, even better: tasks you already did! Go ahead, write it down and immediately cross it off, you deserve it.

\section{The Real Takeaways}

I started this analysis because I thought it would be an incriminating exposé showing how bad everyone is at estimating when they'll complete their work. I was overly aware of the tasks that I copy-pasted for months and worked on for 5x as many hours as planned. I was sure the data would highlight those ``failures''. It turns out, though, that was another false expectation.
%It turns out, though, that expectation was me lying to myself again.

In reality, we did so much good work over the past year. We didn't fail. And even if we hadn't managed to run a single analysis script, we were able to support each other during a ridiculously difficult year and reflect on the things that mattered most to us. 
%When prompted to say what they were proud of, one participant responded ``I was able to do a lot of self care... and learn to slow things down with my physical and mental health; I am managing to still pull through with my goals and being okay with not having perfection."
When asked what we'll remember about these months when looking back ten years from now, one participant responded: ``I realized the importance of health and well-being and put a priority on that for the first time on my life.''

%Sometimes our meetings were more group therapy than group goal planning.
So yes, we only complete some of our planned tasks every week, but also, wow, we complete some of our planned tasks every week! While caring for ourselves and our loved ones! It is difficult to be productive at the best of times, let alone during a global crisis. So if you feel like you are failing because you aren't ``productive enough'', stop lying to yourself.

\section{Acknowledgements}
I want to deeply thank the members of the Weekly Goals Meeting who participated in and encouraged this analysis: Alexander P. Ji, Ivanna Escala, M. Katy Rodriguez Wimberly, Sal Wanying Fu, Anirudh Chiti,  Allen Marquez, and Mimi Truong. Thank you for the supportive accountability over the past year.

\newpage

\appendix

\begin{figure*}[htb!]
\gridline{
	\fig{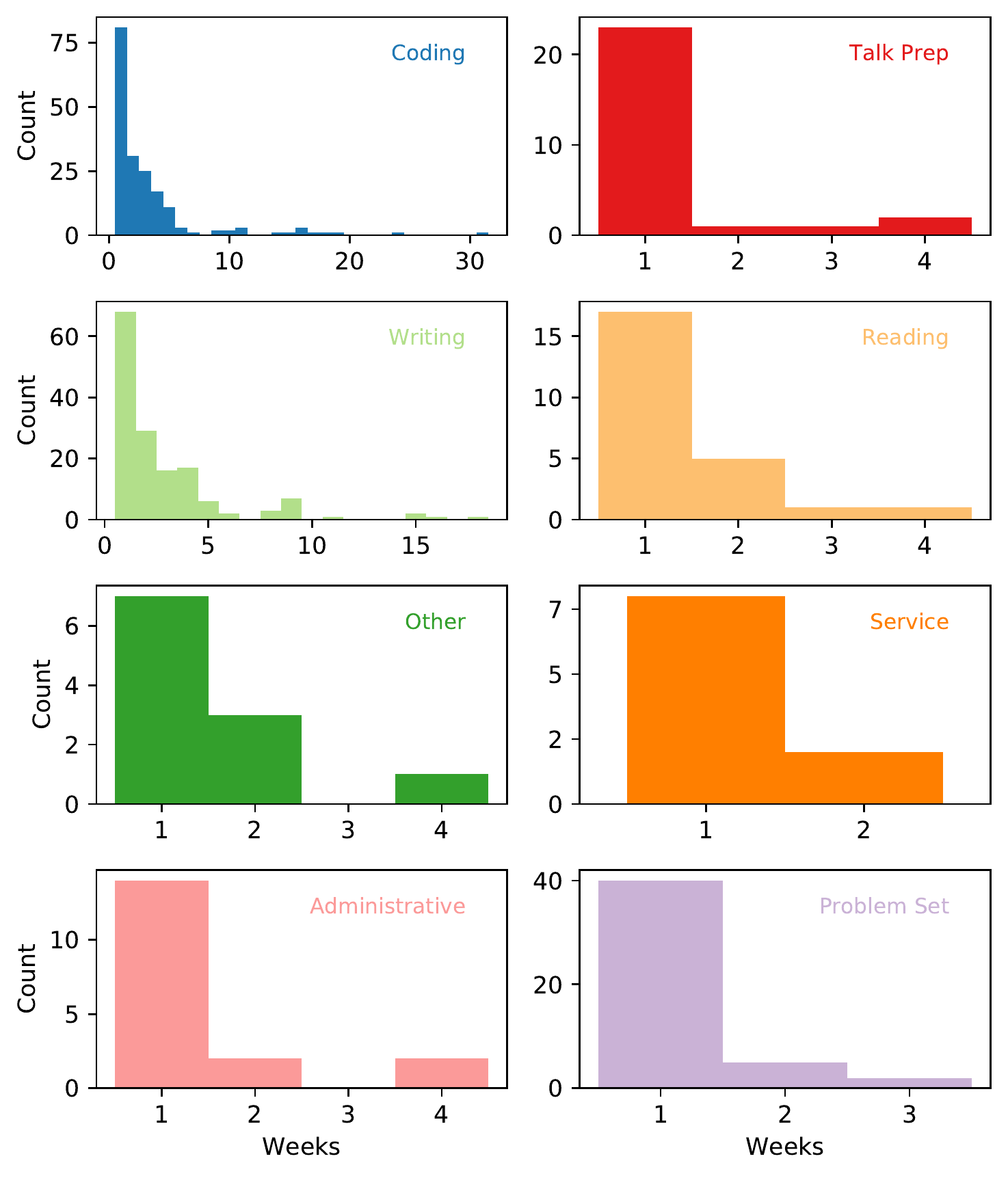}{0.5\textwidth}{(a) How many weeks pass before different types of tasks are completed.}
	\fig{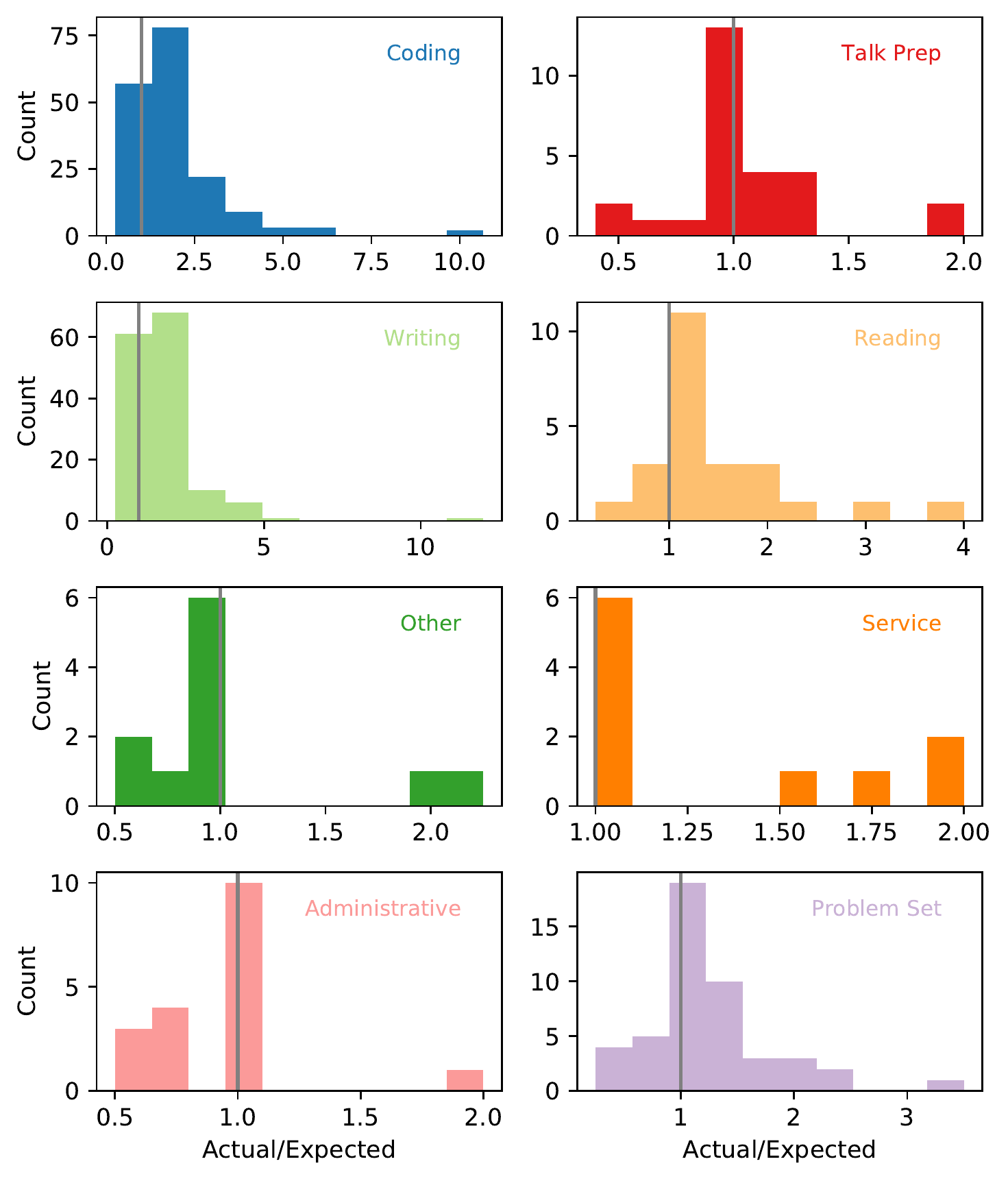}{0.5\textwidth}{(b) How our estimated hours compare to the actual number of hours to complete different tasks.}}
\caption{Histograms showing how quickly different tasks are completed.
\label{fig:hist}}
\end{figure*}

% \section{Appendix information}
% xxx


% \bibliography{sample631}{}
% \bibliographystyle{aasjournal}



\end{document}